# A new self-synchronizing stream cipher

Shihong Wang

*Department of Physics, Beijing Normal University, Beijing 100875, China*
*Science school, Beijing University of Posts and Telecommunications,*
*Beijing 100876 , China*

Huaping Lü

*Department of Physics, Xuzhou Normal University, Xuzhou 221009, China*

Gang Hu

*Department of Physics, Beijing Normal University, Beijing 100875, China*
*Correspondent author (Email: ganghu@bnu.edu.cn)*

## Abstract

A new self-synchronizing stream cipher (SSSC) is proposed based on one-way and nearest neighbor coupled integer maps. Some ideas of spatiotemporal chaos synchronization and chaotic cryptography are applied in this new SSSC system. Several principles of constructing optimal SSSC are discussed, and the methods realizing these principles are specified. This SSSC is compared with several SSSC systems in security by applying chosen-ciphertext attacks. It is shown that our new system can provide SSSC with high security and fairly fast performance.

[1] Manuscript Received January 11, 2005. This work is supported by the National Natural Science Foundation of China under 10335010 and by Nonlinear Science Project.

# I. INTRODUCTION

Cryptography has become one of the fields of modern science since the celebrated Shannon's work [1]. Cryptosystems can be classified to two classes, private key (or say, symmetric key) and public key (unsymmetrical key) ciphers, and the former is dominantly applied when large amounts of data are communicated. Private key cryptosystems can be further classified, according to the different mechanisms of encryption (decryption) and different synchronization structures, to block cipher (BC), synchronous stream cipher (SSC), and self-synchronizing stream cipher (SSSC). So far, in conventional secure communications, some well established BC and SSC ciphers have been popularly used [2, 3], while no SSSC, to our knowledge, has been utilized in commercial services and no SSSC system has been commonly accepted as the standard prototype [4]. The main difficulty of SSSC systems is their low security. The reason is that for these systems the signals transmitted in the public channels serve as the drivings for the synchronization of the receivers, and this structure can be effectively used by intruders to expose much information of the secret key [5]. Nevertheless, SSSC has some remarkable advantages over BC and SSC. It has a property of practical one-time pad if its secret key is well hidden and its periodicity is not reached (an advantage over BC with ECB mode), and it can achieve resynchronization between the transmitter and the receiver by the driving signal after any desynchronization caused by temporal disturbances of communication machines or transmission channel (an advantage over SSC). It is thus a significant task for secure communications to construct desirable SSSC systems which overcome the disadvantages of low security on one hand and keep all the advantages of SSSC on the other hand.

Since the pioneer work of chaos synchronization by Pecora and Carrol, efforts to design optimal SSSC systems are stimulated by chaotic dynamics and chaos synchronization [6–12]. In [13, 14] we suggested to use one-way coupled chaotic maps, incorporating with some simple conventional algebraic operations, to enhance the security of chaotic SSSC; in [15, 16] many space sites of spatiotemporal chaos are used to produce ciphers in parallel and to considerably increase the efficiency of performance; in [16, 17] it is shown that the robustness of secure communication against channel noise can be greatly enhanced by applying SSSC based on spatiotemporal chaos. An experiment of duplex voice communication through Public Switched Telephone Network with embedded CPU set supporting chaotic cryptography is successfully realized [17]. However, in all these papers chaotic cryptography is mainly based on floating-point analytical computation, which is suitable for software implementations, but not convenient for hardware realizations.

In this paper we will go further to apply the idea of one-way coupled map lattice to construct self-synchronizing stream cipher, based entirely on operations of integers. Therefore, the new system, with all advantages of spatiotemporal-chaos-based cryptography well kept, is suited not only for software implementation, but also for hardware uses. The paper is organized as follows. In Sec.II, we specify some conventional SSSC systems appearing in previous papers, and evaluate the security of these systems by chosen-ciphertext attacks. We will show that none of theses systems is secure. In Sec.III we discuss the principles how to construct secure and fast SSSC cryptosystems. The key point for the optimization of both security and performance is that the SSSC should have simple operational algorithms for legal encoders and decoders while complicated canonical transformations for intruders in the same time. Section IV is devoted to describe the structure and the dynamics of our one-way coupled cryptosystem and to explain why this system can reach high security and achieve efficient performance simultaneously. In the last section we make a statistical analyses on the random properties of the output keystream of our system.

## II.  CONVENTIONAL SELF-SYNCHRONIZING STREAM CIPHER SYSTEMS

Since we study cryptography solely in integer number basis, we consider only conventional cryptosystems in comparison. In this section we make general survey of some SSSC systems, and evaluate their securities. For multiple-user secure communications public dynamitic structure is always assumed (Kerckhoff principle). According to the information accessible to the intruder, attacks can be distinguished to ciphertext-only, plaintext-known, chosen-plaintext, and chosen-ciphertext attacks, among which the chosen-ciphertext attacks are the strongest ones, because with these attacks the intruder is assumed to have the maximum information and convenience the legal receiver has except the secret key. The intruder knows the dynamitic structure of the decryption system and can freely use the same legal decoder (i.e., can use the secret key, but cannot see it). In this paper we will use the chosen-ciphertext attacks for evaluation and compare various SSSC systems, including our system suggested in Sec.IV.

We consider the following typical systems for survey and comparison.

### A  Model 1: System with N-order delay piecewise linear combination [5, 18, 19]

$$x_n(1) = c_{n-1}, \ x_n(i) = x_{n-1}(i-1), \ i = 2, 3, ..., N$$

$$k_n = \sum_{i=1}^{N} A_i x_n(i) \mod 2^m \tag{1a}$$

$$c_n = p_n \oplus k_n \ \text{(Encoder)}; \quad p_n = c_n \oplus k_n \ \text{(Decoder)} \tag{1b}$$

where $c_n, p_n$ are ciphertext and plaintext, respectively, and $A_i$, $i = 1, 2, ..., N$ serve as the secret key. Throughout the paper we assume that the variable $x_n(i)$ and $A_i$ has length of $m$ bits, and the secret key $\mathbf{A} = (A_1, ..., A_N)$ has length of $mN$ bits. The algorithm of Eq.(1) is a direct extension of linear feedback shift register and the schemes of encryption is shown in Fig.1(a). In Eq.(1) nonlinearity is introduced by the modulo operations.

In all the following models, the same encryption and decryption transformations (1b) are always implied, and will not be repeated. Equation (1a) will be called as the *operational transformation* which is used by the legal encoder and decoder for cryptographic operations with known secret key. Equation (1a) can be equivalently transformed to the *canonical transformation*

$$k_n = \sum_{i=1}^{N} A_i c_{n-i} \mod 2^m \tag{1c}$$

System (1) possesses some privacy under the ciphertext-only attacks. Its security can be easily broken by using the chosen-ciphertext attacks. Choosing a $N$ ciphertext chain as

$$\mathbf{c}^i = (c_{n-1}, c_{n-2}, ..., c_{n-i}, ..., c_{n-N}) = (0, ..., 0, c_{n-i} = 1, 0, ..., 0) \tag{2}$$

where all ciphertext are zero except $(n-i)$th ciphertext. We immediately have

$$k^i = A_i, \ i = 1, 2, ..., N \tag{3}$$

through Eq.(1c). Any SSSC system obtained from certain linear and nonlinear combinations of Eq.(1) is insecure too, and will not be further considered in this paper.

**B   Model 2: Systems with N-order delay nonlinear combinations of nonlinear maps**

The most general extension of Eq.(1) is a N-order delay with an arbitrary nonlinear output transformation of the keystream

$$k_n = W[x_n(1), ..., x_n(N); \mathbf{A}] \mod 2^m \tag{4a}$$

with $\mathbf{A} = (A_1, ..., A_N)$ being the secret key. High security can be achieved if $W$ has sufficiently complicated operational transformation. However, the complexity of $W$ is restricted by the performance requirement, i.e. in order to reach fast encryption (decryption) speed, the operational transformation $W$ cannot be too complicated. For instance, we can use the transformation of Advanced Encryption Standard (AES) as $W$, then the performance of (4a) must be N times slower than AES for the equivalent security, since in each $W$ operation the SSSC system produces ciphertext $N$ times less than those by AES.

An important characteristic feature of all $N$-order delay systems is that the canonical transformation of SSSC has exactly the same form as the operational transformation (4a)

$$k_n = W[c_{n-1}, ..., c_{n-N}; \mathbf{A}] \mod 2^m \tag{4b}$$

And this is just the very point why SSSC with N-order delay cannot reach high security together with fast performance. This point will be analyzed in detail in the next section. Let us consider a specific form of Eq.(4).

The operational transformation reads

$$x_n(1) = c_{n-1}, \ x_n(i) = x_{n-1}(i-1), \ i = 2, ..., N$$
$$k_n = W[x_n(1), ..., x_n(N); \mathbf{A}] \tag{5a}$$
$$= Q[f(A_1, x_n(1)), ..., f(A_i, x_n(i)), ..., f(A_N, x_n(N))] \mod 2^m$$

leading to the canonical transformation

$$k_n = W[c_{n-1}, ..., c_{n-N}; \mathbf{A}]$$
$$= Q[f(A_1, c_{n-1}), ..., f(A_i, c_{n-i}), ..., f(A_N, c_{n-N})] \mod 2^m \tag{5b}$$
$$= Q(f_1, f_2, ..., f_N) \mod 2^m$$

This is an extension of Eq.(1) by replacing the piecewise linear form of (1) by the general nonlinear form of (5) ($Q$ and $f$ can be nonlinear). The scheme of encryption (5a) is given in Fig.1(b). In order to achieve fast performance the transformations $f$ and $Q$ must not be too complicated. Therefore, we assume in (5) (and in all the following SSSC systems) that function $f$, which is called as the elementary operation, is fairly simple and easily solvable, and function $Q$ (which will be called as the output transformation) itself is also simple and solvable. Namely, if the number of bit equations

$$k^j = Q(f_1^j, ... f_i^j, ..., f_N^j), \tag{6a}$$
$$f_i^j = f(A_i, c_{n-i}^j), \ j = 1, 2, ... \tag{6b}$$

matches the number of unknown bits of $f_i^j$, the variables $f_i^j$ can be easily solved from the function $Q$, and the key bits $A_i, i = 1, 2, ..., N$ can be further solved from $f_i^j$. The security of the whole SSSC system relies on the irreducibility of the couplings of functions $Q$ and $f$. The security is regarded to be weak whenever Eq.(5) can be reduced to separated equations for $Q$ and $f$.

Choosing previous ciphertext chain as $\mathbf{c}^j = (c_{n-1} = 0, ..., c_{n-j} = 1, ..., c_{n-N} = 0)$, and $\mathbf{c}^{i,j} = (c_{n-1} = 0, ..., c_{n-i} = 1, c_{n-i-1} = 0, ..., c_{n-j} = 1, ..., c_{n-N} = 0)$, we can specify $\frac{N(N+1)}{2}$ equations from Eq.(5b) by

$$
\begin{aligned}
k^j &= W(\mathbf{c}^j, \mathbf{A}) \\
&= Q(\phi_1, ..., \psi_j, ..., \phi_N) \\
k^{i,j} &= W(\mathbf{c}^{i,j}, \mathbf{A}) \\
&= Q(\phi_1, ..., \psi_i, ..., \psi_j, ..., \phi_N)
\end{aligned}
\quad (7)
$$

with

$$\phi_j = f(A_j, 0), \ \psi_j = f(A_j, 1) \quad (8)$$

Since Eq.(7) is solvable and for $N \geq 3$ the number of independent equations of Eq.(7) $\frac{N(N+1)}{2}$ is larger than the number $2N$ of unknown variables $(\phi_j, \psi_j, j = 1, ..., N)$ we can obtain the solutions of $\phi_i$ and $\psi_j$, and then further solve $A_j$ from Eq.(8). Therefore, by a suitable chosen-ciphertext attack the task of solving the inverse problem of the general function $W$ can be reduced to the tasks of solving the solvable inverse problems of the functions $Q$ and $f$ separately. This reducibility makes the attack easy, and thus makes system (5) weak in security.

### C  Model 3: N-order tree structure system [20]

The operational system reads

$$x_n(1) = c_{n-1}, \ x_n(i) = x_{n-1}(i-1), \ i = 2, ..., 125$$

$$y_n(j) = f(A_{5j-4}, ..., A_{5j}; x_{n-1}(5j-4), ..., x_{n-1}(5j)), j = 1, ..., 25 \quad (9a)$$

$$z_n(j) = f(A_{125+5j-4}, ..., A_{125+5j}; y_{n-1}(5j-4), ..., y_{n-1}(5j)), j = 1, ..., 5$$

$$k_n = f(A_{151}, ..., A_{155}; z_{n-1}(1), ..., z_{n-1}(5))$$

The canonical form reads

$$k_n = f\{A_{151}, ..., A_{155}; f[A_{126}, ..., A_{130}; f(A_1, ..., A_5; c_{n-1}, ..., c_{n-5}),$$

$$f(A_6, ..., A_{10}; c_{n-6}, ..., c_{n-10}), ..., f(A_{21}, ..., aA_{25}; c_{n-21}, ..., c_{n-25})] \quad (9b)$$

$$, ..., f[A_{146}, ..., A_{150}; f(A_{101}, ..., A_{105}; c_{n-101}, ..., c_{n-105}),$$

$$..., f(A_{121}, ..., A_{125}; c_{n-121}, ..., c_{n-125})]\}$$

This system has chain cryptographic structure shown in Fig.1(c). The output keystream $k_n$ of Eq.(9) contains a single bit, and this makes Eq.(9) flexible in applications.

However, Eq.(9) has still a weakness of reducibility of the key space, and the secret key bits can be solved part by part in the equations by suitably chosen ciphertexts. The costs of chosen-ciphertext attacks can be much lower than the brute force attack in the secret key space. For instance, we choose a set of ciphertexts **c** with arbitrarily fixing all ciphertext bits $c_{n-i}$, $6 \leq i \leq 125$ and changing the first five bits $c_{n-i}$, $1 \leq i \leq 5$ only. The first five bits may be grouped as

$$
\begin{aligned}
&\mathbf{c}^1 && (0,0,0,0,0) \\
&\mathbf{c}^{2-6} && (1,0,0,0,0), (0,1,0,0,0), ... \\
&\mathbf{c}^{7-16} && (1,1,0,0,0), (1,0,1,0,0), ... \\
&\mathbf{c}^{17-26} && (1,1,1,0,0), (1,0,1,1,0), ... \\
&\mathbf{c}^{27-31} && (1,1,1,1,0), (1,0,1,1,1), ... \\
&\mathbf{c}^{32} && (1,1,1,1,1)
\end{aligned}
\tag{10}
$$

Suppose we observe $k^j = 0$ for a group of $\mathbf{c}^j$'s and $k^{\hat{j}} = 1$ for the other $\mathbf{c}^{\hat{j}}$'s. These data determine two groups of equations in the first level

$$
\begin{aligned}
y^j(1) = f(A_1, ..., A_5; \mathbf{c}^j) = \begin{cases} 0 \\ 1 \end{cases}, \\
y^{\hat{j}}(1) = f(A_1, ..., A_5; \mathbf{c}^{\hat{j}}) = \begin{cases} 1 \\ 0 \end{cases}
\end{aligned}
\tag{11}
$$

From which the part of secret key $A_1, ..., A_5$ can be solved with $\frac{1}{2}$ uncertainty. The cost of solving all the 125 key bits of the first level is limited by the exhaustive searching of $2^{25}$ tests (in Eq.(11) one set is correct among the two possible sets of solutions). With the similar procedure we can solve the other secret key. Then the same approach can be applied to key bits of the second and third levels.

The weakness of Eqs.(5) and (9) is clearly due to the reducibility of the secret key space, which considerably reduces the computational difficulty in exposing the secret key. The same problem exists also in other SSSC systems. For instance, in [20] the author suggested SSSC system with bit diffusion much stronger than system (9), but the system is still not strong enough to entirely exclude the similar reducibility problem and it is weak against the chosen-ciphertext attacks.

## III. PRINCIPLES TO CONSTRUCT EFFECTIVE SSSC

Before coming to the discussion of principles, let us first consider why all the models in Sec. II have difficulties to achieve high security. We analyze this problem by taking the example of the SSSC system Eq.(4). The encoder and decoder use the operational transformation Eq.(4a) to perform cryptography. The transformation Eq.(4a) cannot be too complicated due to the requirement of fast encryption (decryption) speed. In this regard, we simply assume that if all the dynamic variables $x_n(i)$, $i = 1, 2, ..., N$ are known to the public, the intruder can expose the secret key $\mathbf{A}$ with no essential difficulty (i.e., the system is weak for the security). Nevertheless, $x_n(i)$ are not necessarily known for the intruder. What is definitely open to the intruder is the canonical transformation Eq.(4b) and arbitrarily many data of $c_n$ in the equation. Unfortunately, with the N-order delay form of Eq.(4), the canonical transformation is exactly the same as the operational transformation with the identities of $c_{n-i} = x_n(i)$. For Eq.(4), chosen-ciphertext means chosen-dynamical variables, and this makes Eq.(4) weak in security unless the output function $W$ takas complicated form which requires much computational cost. Thus, the two important properties for any cryptosystem, high security and fast performance, cannot be realized simultaneously for the SSSC systems of models 1-3, and for all SSSC models, known so far.

One hopeful way out the above difficulty is to design SSSC systems which have canonical transformation dramatically different from their corresponding operational transformation. Due to the characteristic of driving synchronization, any SSSC system must have the following general structure

$$k_n = W(\mathbf{A}; c_{n-i_1}, c_{n-i_2}, ..., c_{n-i_q}) \qquad (12)$$

with $\mathbf{A}$ being the secret key, and function $W$ open to the public. Therefore, the encoder (decoder) is fully observable, and the central focus of construction of SSSC systems is to make the function $W$ as complicated as possible. And on the other hand, the corresponding operational transformation should be made as simple as possible to allow the legal encoder and decoder to perform fast cryptography.

With the available canonical form of Eq.(12), the chosen-ciphertext attacks are particularly strong, much stronger than plaintext-known and chosen-plaintext attacks. Because, properly choosing ciphertext sequences $c= (c_{n-i_1}, c_{n-i_2}, ..., c_{n-i_q})$ may effectively expose the weakest points of the transformation $W$. Therefore, we can briefly summarize the following conditions for designing SSSC systems resistant against chosen-ciphertext attacks:

(i) In Eq.(12) any change of any input bit of $c_{n-i_j}, j = 1, 2, ..., q$, can influence any output bit of $k_n$.

(ii) In the response of any $k_n$ bit on the change of any $c_{n-i_j}$ bit, all key bits of $\mathbf{A}$ must be fully involved as a whole. Namely, $\mathbf{A}$ is irreducible in the sense: one cannot solve any partial bits of $\mathbf{A}$ independent of (or weakly dependent on) other $\mathbf{A}$ bits through Eq.(12) with properly selected $c_{n-i_j}$ patterns.

It can be easily checked that condition (i) is weak, which is fulfilled by all models in Sec. II. However, condition (ii) is much stronger, and all models in Sec. II do not meet this condition.

(iii) Conditions (i) and (ii) are necessary but not sufficient for optimal SSSC. A further requirement is: The strong confusions and diffusions of (i) and (ii) must be made effective (i.e., with low computational cost of the legal encoder and decoder), namely, the complicated canonical transformation must be provided by relatively simple operational transformation.

Let us now discuss how to construct an optimal SSSC system based on the above principles. A reasonable improvement from Eq.(4a) is to change its N-order delay by introducing couplings between different maps, and the simplest coupling is the one-way nearest neighbor coupling

$$\text{Input function } x_n(0) = f_0(\mathbf{A}, c_{n-1}) \tag{13a}$$

$$\text{Cryptographic dynamics } x_n(i) = f[A_i, x_{n-1}(i), x_{n-1}(i-1)], \ i = 1, ..., N \tag{13b}$$

$$\text{Output function } \ k_n = Q[\mathbf{A}, x_n(1), x_n(2), ..., x_n(N)] \tag{13c}$$

The elementary operational function $f$ and output function $Q$ must be fairly simple in the sense that $A_i$ can be easily solved if enough variable data $x_n(i), x_{n-1}(i)$ and $x_{n-1}(i-1)$ are available. The security of system (13) relies mainly on the nonlinearity of $f$ and the coupling induced irreducibility between different maps.

It is emphasized that with Eq.(13) the synchronization between the encoder and decoder is not necessarily reached, i.e., it is possible that the decoder

$$y_n(0) = f_0(\mathbf{A}, c_{n-1}) \tag{14}$$
$$y_n(i) = f[A_i, y_{n-1}(i), y_{n-1}(i-1)], \ i = 1, 2, ..., N$$

may not achieve $\mathbf{y}_n = \mathbf{x}_n$, as time goes on, from an arbitrary initial condition (different from the initial condition of the encoder). In order to guarantee synchronization, the functions $f[A_i, y_{n-1}(i), y_{n-1}(i-1)]$ must have certain forgetting mechanism for the initial value of $y_0(i)$. Suppose $f$ forget the initial condition after $\nu$ iterations, the total synchronization time of the decoder is

$$T_t = N\nu \tag{15}$$

We adopt in the following discussion

$$\nu = 2 \tag{16}$$

and the forgetting mechanism can be thus represented by

$$x_n(i) = f[A_i, x_{n-1}(i), x_{n-1}(i-1)] \tag{17a}$$
$$x_n(i) = f[A_i, f(A_i, x_{n-2}(i), x_{n-2}(i-1)), x_{n-1}(i-1)] \tag{17b}$$
$$= \psi[A_i, x_{n-1}(i-1), x_{n-2}(i-1)] \tag{17c}$$

The neglection of $x_{n-2}(i)$ in (17c) is guaranteed by the two-step forgetting law, while the derivation from $f$ to $\psi$ is model dependence. With Eq.(17) we have synchronization time

$$T_t = 2N \tag{18}$$

In Sec.II, for all models investigated we give explicitly both operational forms Eqs.(1a), (4a), (5a), and (9a), and their corresponding canonical forms of Eqs.(1c), (4b), (5b), and (9b). The operational forms are used by the encoder (also decoder) to perform encryption (decryption) with known secret key **A**, while the canonical ones can be used by the intruder to break the security of the systems, i.e., to expose the secret key **A** from the publicly accessible transmitted ciphertext. Therefore, all wisdom in constructing optimal SSSC systems is to complicate the latter as much as possible on one hand, while simplify the former as much as possible on the other hand, for achieving high security and fast performance simultaneously.

The operational form Eq.(13) is similar to Eq.(5a) with its computational cost if we take the same system size $N$ and similar elementary operations $f$ and output function $Q$. However, the canonical form corresponding to (13) is no longer the same as Eq.(17). In the next section we will find that this canonical form, specified with given $f$ and $Q$, is extremely complicated, and incomparably more complicated than the canonical forms of all the previous models with the simple complexity of the operational form.

## IV. A NEW SELF-SYNCHRONIZING STREAM CIPHER MODEL

Now we specify all the details of system (17). The number of block bits of a single map $m$ and the system size $N$ are taken as

$$m = 32, \quad N = 6 \tag{19a}$$

respectively. Therefore, we have the number of variable bits

$$J = 192 \tag{19b}$$

which leads to average period

$$T_p \approx 2^{96} \tag{19c}$$

if the transformation (17) is practically random. Consistently, we use a secret key of 128 bits $A_1, A_2, A_3, A_4$, which can be expanded to $A = (A_1, A_2, ..., A_8)$ as

$$A_5 = E[(A_1 \oplus A_2 \oplus A_3 \oplus A_4) \lll 13] \tag{20}$$

$$A_6 = E[(A_2 \oplus A_3 \oplus A_4 \oplus A_5) \lll 13]$$

$$A_7 = E[(A_3 \oplus A_4 \oplus A_5 \oplus A_6) \lll 13]$$

$$A_8 = E[(A_4 \oplus A_5 \oplus A_6 \oplus A_7) \lll 13]$$

with each having length of 32 bits. The operation $x \lll y$ stands for a left cycleshift of $x$ by $y$ bits, and $\oplus$ means bitwise XOR. $E$ is a nonlinear operation.

For simplicity we set the input function as an identity operator $x_n(0) = c_n$. The elementary operator $f$ (shown in Fig.1(d)) is defined by three suboperations $E$, $F$, and $G$ as

$$\begin{aligned} x_n(i) &= f[A_i, x_{n-1}(i), x_{n-1}(i-1)] \\ &= E[A_i \oplus G x_{n-1}(i) \oplus F x_{n-1}(i-1)] \\ &= E_i[G x_{n-1}(i) \oplus F x_{n-1}(i-1)], \ i = 1, 2, ..., 6 \end{aligned} \tag{21a}$$

The output function is defined as

$$\begin{aligned} k_n &= Q[x_n(N), x_{n-1}(N-1)] \\ &= E[x_n(N) \oplus (x_{n-1}(N-1) \lll 5) \oplus A_{N+1}] \oplus A_{N+2} \end{aligned} \tag{21b}$$

Operators $G$, $F$ and $E_i$ are defined as

$$Gx = x >> 16, \tag{22a}$$

$$Fx = x \oplus (x \ggg 11) \oplus (x \lll 11)$$

$$E_i x = E(x \oplus A_i), \ i = 1, 2, ..., 6$$

In Eq.(22a) the operation $x >> y$ denotes a right shift of $x$ by $y$ bits. The operation $E$ is defined by four independent and identical $8 \times 8$ bit S-box nonlinear transformations. First, we split a 32-bits variable $x$ into four bytes $x = x_1 | x_2 | x_3 | x_4$. Each byte takes a nonlinear map from 8 bits to 8 bits. Finally, four bytes are combined to a new word $E(x) = sbox(x_1) | sbox(x_1) | sbox(x_3) | sbox(x_4)$. Three possible choices of S-box are given as:

(i) Quadratic congruential transformation

$$f(x) = 3x^2 + 2x + 1 \mod 256 \tag{22b}$$

(ii) Multiplicative inverse transformation

Let polynomials $a(x)$ and $b(x)$ be the input and output elements of S-box. Given a irreducible polynomial $p(x) = x^8 + x^5 + x^3 + x + 1$, the S-box is defined to be

$$a(x) \cdot b(x) = 1 \mod p(x) \tag{22c}$$

0 element is mapped to itself.

(iii) Random map

Given $i = 0, 1, ..., 255$. S-box is defined as a one-to-one random map

$$j = S(i) \tag{22d}$$

It is emphasized that all the elementary operations $G$, $F$ and $E$ are common and simple. They are equally used in Eqs.(5) and (13). These operations cannot make model 2 secure. However, together with the coupling structure these operations make Eq.(13) highly secure with computational expense no more than Model 2.

We can now derive the canonical transformation corresponding to the operational transformations (21). Let us start from the first map

$$x_n(1) = E_1[Gx_{n-1}(1) \oplus Fx_{n-1}(0)]$$
$$= E_1\{GE_1[Gx_{n-2}(1) \oplus Fx_{n-2}(0)] \oplus Fx_{n-1}(0)\} \quad (23a)$$
$$= E_1[GE_1Fx_{n-2}(0) \oplus Fx_{n-1}(0)]$$

Going further to $x_n(2)$, we have

$$x_n(2) = E_2[GE_2Fx_{n-2}(1) \oplus Fx_{n-1}(1)]$$
$$= E_2\{GE_2FE_1[GE_1Fx_{n-4}(0) \oplus Fx_{n-3}(0)] \quad (23b)$$
$$\oplus FE_1[GE_1Fx_{n-3}(0) \oplus Fx_{n-2}(0)]\}$$

From Eqs.(23a) and (23b) it is easy to derive an iteration relation. Given

$$x_n(j) = P[x_{n-j}(0), x_{n-j-1}(0), ..., x_{n-2j}(0)]$$

we have

$$x_n(j+1) = E_{j+1}[GE_{j+1}FD^2P \oplus FDP] \quad (23c)$$

with $D$ being a one-step time delay operator

$$DP(x_{n-j}(0), x_{n-j-1}(0), ..., x_{n-2j}(0)) = P(x_{n-j-1}(0), x_{n-j-2}(0), ..., x_{n-2j-1}(0)) \quad (23d)$$

Step by step, the general dynamic variable $x_n(i)$ can be written in terms of $c_n$ as

$$x_n(i) = \prod_{j=1}^{i}{'} E_j(GE_jFD^2 \oplus FD)c_n \quad (24)$$

where $\Pi'$ denotes an ordered operator product $E_i(GE_iFD^2 \oplus FD)E_{i-1}(GE_{i-1}FD^2 \oplus FD)...E_1(GE_1FD^2 \oplus FD)$.

Finally, the canonical transformation can be specified analytically, based on Eqs.(21) and (24), as

$$k_n = Q[x_n(N), x_{n-1}(N-1)] \tag{25}$$
$$= Q[\prod_{j=1}^{N}{}' E_j(GE_jFD^2 \oplus FD)c_n, \prod_{j=1}^{N-1}{}' E_j(GE_jFD^2 \oplus FD)c_{n-1}]$$

At this point a comparison of our model with the models in Sec. II is in order. First, the operational transformation is simple, at least, equivalently simple as model Eq.(5a). This simplicity allows fast performance speed. However, the operational transformation of Eq.(21) is as weak as the models in Sec. II in the sense that the secret key can be easily exposed if arbitrarily many variable data $x_n(i)$ are known. This weakness causes the low security of all models in Sec. II where the canonical transformations, which are used by the intruder to make cryptoanalysis, have exactly the same forms as the corresponding operational forms [with available ciphertext data $c_n$ the intruder knows all data of $x_n(i)$]. However, this problem does not exist for our model where the canonical form Eq.(25) is considerably different from the operational form Eq.(21). Therefore, with the available data $c_n$ the variable values $x_n(i)$ are not directly accessible. Complicated calculation is needed for computing $x_n(i)$ from $c_n$. We have indeed design a SSSC system with a simple operational transformation for easy computations of the legal encoder and decoder and a complicated canonical transformation for difficult analysis of the illegal intruder, i.e., we achieve fast cryptographic performance and high security simultaneously.

Comparing Eq.(25) with Eq.(5b), the canonical transform Eq.(25) has mixed both combination and composition transformations in a complicated way no any transformation is reducible from others. This therefore very much strengthens the security of the SSSC system. A remarkable point is that this complication is not realized by complicating the operational transformation, but by introducing a memory mechanism Eq.(17), and this mechanism does not slightly increase the encryption (decryption) cost.

Though formula $W$ in Eq.(25) can be explicitly derived in an operator form, the actual function is extremely complicated. Even as we take the simplest quadratic S-box Eq.(22b) for nonlinear operator $E$, $W$ contains $A_1$, $A_2$, $A_3$, $A_4$ polynomials with powers up to $2^{13} \approx 8000$ in Eq.(21). With the bit shift, cycle-shift and XOR bitwise operations, the solutions of algebraic equations have to be obtained on the bit level. With 128 key bits and $2^{13}$ power, the computational cost of solving Eq.(25) from $k_n$ to 128-bit key is huge. It is hardly possible to break the security of system (13) by analytical calculations with cost less than exhaustive searching in $2^{128}$ key space.

The complicated $W$ formula of Eq.(25) not only makes attacks based on analytical computation difficult, but also guarantees strong bit confusion and diffusion rates, and make attacks based on statistical analysis and key sensitivity analysis difficult too. These points will be discussed in our future paper in detail.

By adjusting parameters $N$ and the length of the secret key, we can flexibly meet different requirements of practical communications. System of (21) with $N = 6$ (192 bits variable) and $M = 4$ (128 bits secret key) is estimated to have similar security level as AES. And Eq.(21) with $N = 4$, $M = 2$ (128 bits of variable space; 64 bits of secret key) may be more secure than DES. It is remarkable that with $N = 6$ the cryptographic speed of Eq.(21) is similar to that of AES with 128 bits block and 128 bits key. Nevertheless, there is no any proof on the security estimations. Detailed and convincing comparisons our new SSSC system with other conventional and chaotic cryptosystems are tasks of our forthcoming works.

## V. STATISTICAL ANALYSES OF KEYSTREAM SEQUENCES

In this closing section we study the random properties of the keystream of our new SSSC system, applying the NIST statistical Test Suite [21] which includes 16 statistical tests to demonstrate various statistical properties of data. For a fixed binary sequence statistical tests produce relevant $P_{-value}$, and analysis of $P_{-value}$ indicates a deviation from pure randomness. In order to value the $P_{-value}$, NIST [21] supplies two approaches, the examination of the proportion of sequences that pass a statistical test and the checking of the uniformness of the $P_{-value}$ distributions. We now consider the first examination. If $T$ binary sequences were tested, $T_e$ binary sequence had $P_{-value} \geq \alpha$ (choosing the significance level $\alpha = 0.01$), then the passing proportion is $\eta = \frac{T_e}{T}$. The confidence interval of this binomial distribution is defined as [21]

$$\eta_c = (1 - \alpha) \pm 3\sqrt{\frac{\alpha(1-\alpha)}{T}} \qquad (26)$$

If the proportion $\eta$ falls inside of this interval, then the keystream is regarded to successfully pass the random checking. In our computation we take $T = 1024$, and then confidence interval $\eta_c = [0.98067, 0.99933]$. Figure 2(a) illustrates the passing proportions corresponding 42 $P_{-value}$'s for 16 statistical tests, and we can see that the keystream well pass all the statistical tests.

To check the uniformness of the distributions of the $P_{-value}$, we take $\chi^2$ test. $\chi^2$ is defined as

$$\chi^2 = \sum_{i=1}^{10} \frac{(T_i - \frac{T'}{10})^2}{\frac{T'}{10}}, \quad T' = \sum_{i=1}^{10} T_i \qquad (27)$$

where $T'$ is the sum of all $P_{-value}$, and $T_i$ is the number of $P_{-value}$ in sub-interval $i$. Choosing significance level $\alpha = 0.01$ and $T' = 400$, if $\chi^2 \leqslant \chi_\alpha^2(9) = 21.666$, the distribution of $P_{-value}$ is regarded to be sufficiently uniform. Figure 2(b) illustrates $\chi^2$ values corresponding 42 $P_{-value}$ and shows all $\chi^2 \leqslant \chi_\alpha^2(9)$.

## REFERENCES

[1] C. E. Shannon, "Communication theory of secrecy systems", Bell Syst. Tech. J., vol. 28, pp. 656-715, 1949.

[2] NIST, "Report on the Development Advanced Encryption Standard (AES)," [online]. Available: http://csrc.nist.gov/encryption/aes, Oct. 2000.

[3] Chapter 3, [on line] Available: http://www.RSAsecurity.com/rsalabs/crypto FAQ.

[4] NESSIE IST-1999-12324, "Performance of Optimized Implementations of the NESSIE

Primitives," [online]. Available: http://www.cosic.esat.kuleuven.ac.be/nessie, Oct. 2002.

[5] F. Dachselt, and W. Schwarz, "chaos and cryptography", IEEE trans. Circuits syst. I, vol. 48, no. 12, pp. 1498-1509, 2001.

[6] L. M. Pecora and T. L. Carroll, "Synchronization in chaotic systems," Phys. Rev. Lett., vol. 64, no. 8, pp. 821-824, 1990.

[7] L. M. Cuomo and A. V. Oppenheim, "Circuit implementation of synchronized chaos with applications to communications,", Phys. Rev. Lett., vol. 71, pp. 65-69, 1993.

[8] L. Kocarev, K. S. Halle, K. Eckert, L. O. Chua and U. Parlitz, "Experimental demonstration of secure communication via chaotic synchronization," Int. J. Bif. and chaos, vol. 2, no. 3, pp. 709-713, 1992.

[9] L. Kocarev and U. Parlitz, "General Approach for Chaotic Synchronization with Applications to Communication," Phys. Rev. Lett., vol. 74, no. 25, pp. 5028-5031, 1995.

[10] J. H.Xiao, G. Hu and Zh. L. Qu, "Synchronization of Spatiotemporal Chaos and Its Application to Multichannel Spread-Spectrum Communication," Phys. Rev. Lett., vol. 77, pp. 4162-4165, 1996.

[11] S. Sundar and A. A. Minai, "Synchronization of Randomly Multiplexed Chaotic Systems with Application to Communication," Phys. Rev. Lett., vol. 85, no. 25, pp. 5456-5459, 2000.

[12] J. Garcia-Ojalvo and R. Roy, "Parallel communication with optical spatiotemporal chaos," IEEE trans. Circuits syst. I, vol. 48, no. 12, pp. 1491-1497, 2001.

[13] S.H. Wang, J.Y. Kuang, J.H. Li, Y.L. Luo, H.P. Lu and G. Hu, "Chaos-based secure communications in a large community," Phys. Rev. E, vol. 66, pp. 065202, 2002.

[14] G.N. Tang, S.H. Wang, H.P. Lu, G. Hu, "Chaos-based cryptograph incorporated with S-box algebraic operation," Phys. Lett. A, vol. 318, pp. 388-389, 2003.

[15] S.H. Wang, W.P. Ye, H.P. Lu, J.Y. Kuang, J.H. Li, Y.L. Luo, and G. Hu, "Spatiotemporal-chaos-based Encryption Having Overall Properties Considerably Better Than Advanced Encryption Standard", Commu. Theor. Phys. vol. 40, no. 1, pp.

57-61 , 2003.

[16] H.P Lü, et al, "A new spatiotemporally chaotic cryptosystem and its security and performance analyses," Chaos, vol. 14, no. 3, pp. 617-629, 2004.

[17] W.P. Ye, et al, "Experimental realization of a highly secure chaos communication under strong channel noise," Phys. Lett. A, vol. 330, no. 1-2, pp. 75-84, 2004.

[18] M. Gotz, K. Kelber, W. Schwarz, "Discret-time chaotic encryption systems– part I: statistical design approach," IEEE trans. Circuits syst. I, vol. 44, no. 10, pp. 963-970, 1997.

[19] W. J. Savage, "self-synchronising scrambler for data transmission," Bell. Syst. Tech., pp. 449-487, Feb., 1967.

[20] G. J. Kuhn, "Algorithms for self-synchronising ciphers", Communications and Signal Processing, Proceedings., COMSIG 88., pp. 159-164, 1988.

[21] NIST, "A statistical test suite for random and pseudorandom number generators for cryptographic applications," [online] Available: http://csrc.nist.gov/encryption/, May, 2001.

# Figure Captions

Fig.1 Encryption schemes of Eq.(1) (a); Eq.(5) (b); Eq.(9) (c); and Eq.(21a) (d), respectively.

Fig.2 The statistical tests of the keystream of the SSSC system of Eq.(21). Successful passes are observed in all tests. (a) Proportions of sequences successfully pass all the 16 tests (corresponding 42 $P_{-value}$'s). The plots are well between the lower threshold 0.98067 and the upper one 0.99933. (b) $\chi^2$ values defined in Eq.(27) for the 42 $P_{-value}$. $\chi^2 \leq \chi^2_{0.01}(9) = 21.666$ (dot lines).

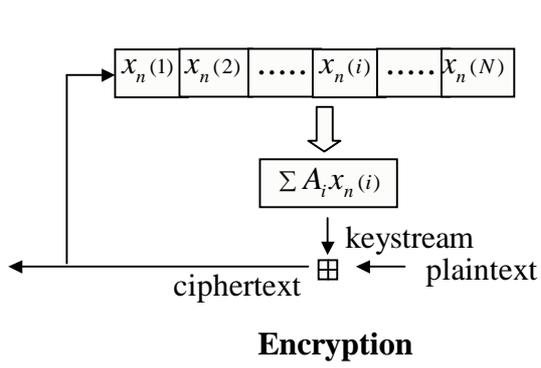
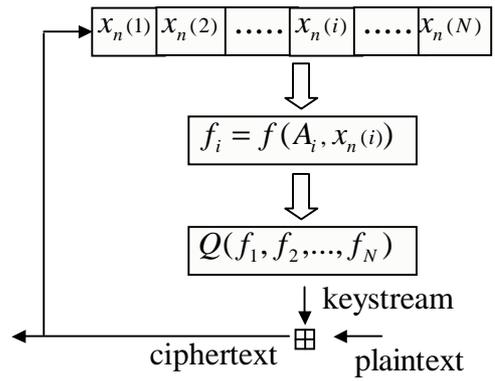

Fig. 1(a)  Fig. 1(b)

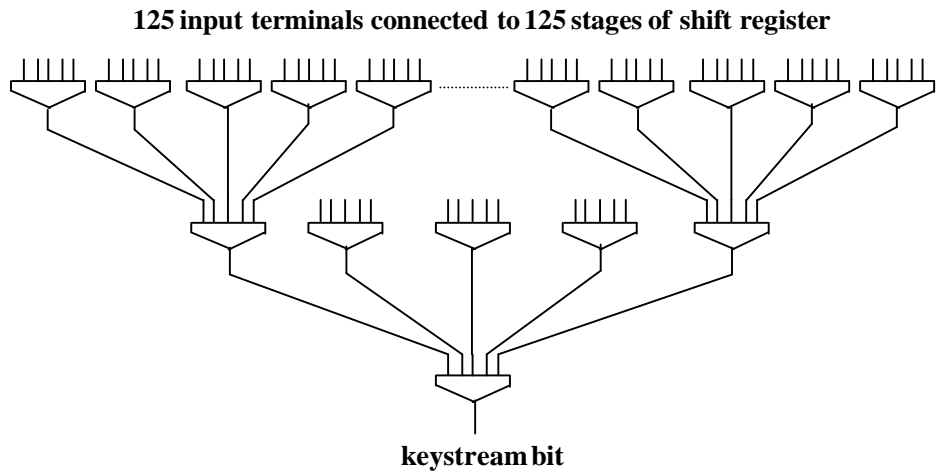

Fig. 1(c)

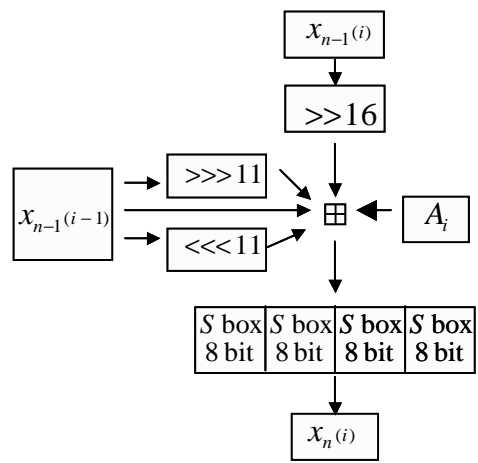

Fig. 1(d)

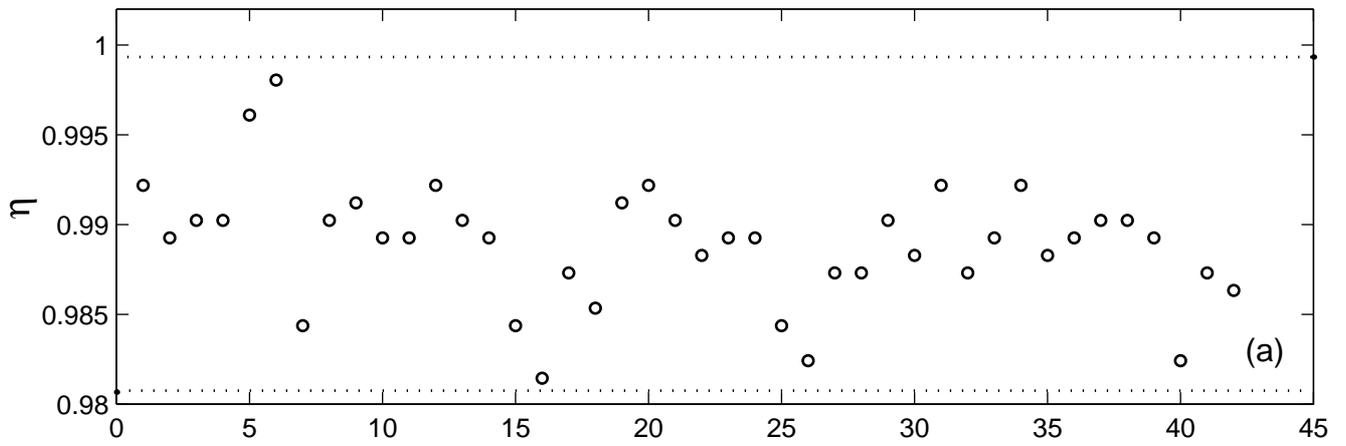

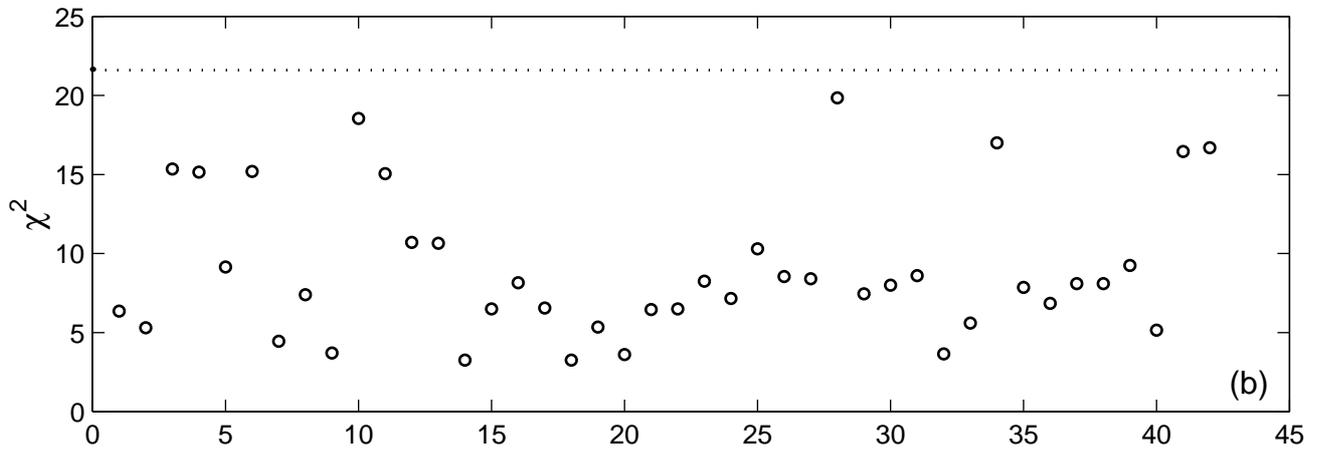

Fig.2